\documentclass{ws-rv9x6}
\usepackage{ws-rv-van} 
\begin{document}
\chapter{Towards noncommutative gravity}
\author[D. V. Vassilevich]{Dmitri Vassilevich}

\address{CMCC, Universidade Federal do ABC, Santo Andr\'e, SP, Brazil\\
Department of Theoretical Physics, 
St.Petersburg University, Russia}
\begin{abstract}
In this short article accessible for non-experts I discuss possible ways
of constructing a non-commutative gravity paying special attention
to possibilities of realizing the full diffeomorphism symmetry
and to relations with $2D$ gravities.
\end{abstract}

\body

\section{Preliminaries}
For the first time I met Wolfgang Kummer in 1992. It happened on
my way back from Italy to St.Petersburg. At that time, a hundred
of US dollars was a fortune in Russia. Therefore, to save money I took
a train going through Vienna, and not a plane flying over it.
The most natural decision was to stop in Vienna for a couple of
days and give a seminar at TU. This is how one of the most fruitful
and exciting collaborations in my life started, and this is also
a very rare example of a positive effect of severe financial difficulties.

The Vienna School of 2D gravity was an amazingly successful project,
see \cite{Bergamin:2008ka}. To keep it running, new interesting directions
of research were always needed. About 2005 I told Wolfgang about 
my recent work 
on noncommutative (NC) gravity in two dimensions \cite{Vassilevich:2004ym}
which almost literally repeated some of the steps done previously in the
commutative case. We decided to return to this after completing our current
work. Unfortunately, deteriorating health did not allow Wolfgang to take up 
this job. This short article is a kind of a proposal for a ``Vienna-style''
NC gravity. This is not a (mini)review, with most visible consequence
that the literature is incomplete. I am asking all authors whose papers
will not be mentioned for understanding. For a systematic overview of NC
gravities the reader may consult the paper by Szabo\cite{Szabo:2006wx}.

Generally speaking, the desire to construct an NC gravity is very natural.
One of the main arguments in favor of noncommutativity comes from 
gravity\cite{Doplicher:1994tu}. Particular ways to realize noncommutativity
differ much from model to model.

To stay closer to Vienna, whenever possible, I will discuss noncommutative
counterparts of dilaton gravities in two dimensions 
(see Ref.\cite{Grumiller:2002nm} for a review). In the commutative case, the
classical first-order action reads
\begin{equation}
S=\int_M \left[ X_aDe^a +Xd\omega +\epsilon \left(U(X)X^aX_a/2 +V(X)
\right)\right]\,,\label{1st}
\end{equation}
where $a=0,1$ is a Lorentz index, $e^a$ and $\omega$ are the zweibein
and connection one-forms respectively, $\epsilon$ is a volume two-form,
$X$ is the dilaton, and $X^a$ is an auxiliary field which generates the
torsion constraint. $De^a=de^a+\varepsilon^a_{\ b}\omega\wedge e^b$,
where $\varepsilon^{ab}$ is the Christoffel symbol. $U(X)$ and $V(X)$
are two arbitrary functions called the dilaton potentials.
With the choice $U(X)=0$, $V(X)\propto X$ one obtains the 
Jackiw-Teitelboim model\cite{JT}. Other choices reproduce all
gravity models in two dimensions, see Ref.\cite{Grumiller:2006rc}.

\section{What can we call a noncommutative gravity?}
In principle, any theory containing some effects of noncommutativity
of the coordinates and looking more or less like a gravity theory
may be called a noncommutative gravity. The problem is that the people
working on a particular approach are (naturally) more enthusiastic 
about it than the rest of the community. Therefore, I asked myself,
what kind of noncommutative gravity theory could have a chance to
satisfy Wolfgang? An answer to this question seems to be a rather
strict point of view on NC gravity. 

To construct a gravity one first needs a manifold. NC manifolds may be
understood through the Gelfand-Naimark duality. To a manifold $M$ 
one can associate
a commutative associative algebra $C^\infty (M)$ of smooth functions.
Under certain restrictions, each commutative associative algebra is
an algebra of smooth functions on some manifold. In this sense, an
algebra $A$, which is a noncommutative associative deformation
of $C^\infty (M)$ defines an NC deformation of $M$. Most conveniently
the deformation is done by replacing the point-wise product $f_1\cdot f_2$
by a noncommutative star product $f_1\star f_2$, which can be presented
as
\begin{equation}
f_1\star f_2 =f_1 \cdot f_2 +\frac i2 \theta^{\mu\nu}(x)
\partial_\mu f_1 \cdot \partial_\nu f_2 +O(\theta^2)\,.
\label{star}
\end{equation}
Because of the associativity, $\theta^{\mu\nu}$ is a Poisson bivector, i.e.
it has to satisfy the Jacobi identity. Note, that in two dimensions the 
Jacobi identity is satisfied any antisymmetric tensor $\theta^{\mu\nu}(x)$.

For a constant $\theta$ there exists a simple (Moyal) formula
for the star product
\begin{equation}
(f_1\star_M f_2)(x)=\exp\left( \frac i2 \theta^{\mu\nu}
\partial^x_\mu\partial_\nu^y \right)f_1(x)f_2(y)\vert_{y=x}\,.
\label{Moyal}
\end{equation}

Next, one has to satisfy the relativity principle, i.e., one 
should realize the group of diffeomorphisms (or a deformation of this
group) on an NC manifold. Then one has to construct invariants
which in the commutative limit $\theta\to 0$ reproduce the 
Einstein-Hilbert action coupled to matter fields. This program, upon
completion, should give an NC gravity.

None of the existing approaches to the NC gravity fulfills strictly
all the requirements formulated above, but we still can learn a lot
from each of them.

\subsection{Minimalistic approaches}
These are approaches which are not even trying to construct a full
NC gravity but instead focus on some selected features of NC theories.
For example,
in one of such approaches, reviewed in Ref.\cite{Nicolini:2008aj},
the nonlocality, which is a characteristic feature of NC theories,
is modelled by delocalization of sources in otherwise commutative
theories. Such approaches are very useful in one wishes to understand
what kind of physical effects may follow from the
noncommutativity, but they are
not designed to check theoretical consistency.

\subsection{Seiberg-Witten map}
In 1999 Seiberg and Witten \cite{Seiberg:1999vs} discovered a map
between commutative and noncommutative gauge theories. Due to this
map, gauge symmetries, including diffeomorphisms, can be realized
by standard commutative transformations on commutative fields. The
NC fields are expressed through power series in $\theta$ with
growing number of commutative fields and their derivatives. This
map was applied also to gravity, and even some physical effects
were studied, see e.g. \cite{SWgra}. With higher orders of
$\theta$ technical difficulties in applying the Seiberg-Witten map
grow fast, so that no one was able to go beyond the second order.
Because of this, this method can hardly be considered as an
ultimate solution of the problem of constructing an NC gravity,
but it gives a very valuable information: the statement that such
a theory does exist at least in the form of power series.

\subsection{Gauging symplectic diffeomorphisms}
Looking at the formula (\ref{Moyal}) one immediately sees a source
of the problems with the diffeomorphisms: $\theta^{\mu\nu}$ looks
as a tensor, but the formula (\ref{Moyal}) is not tensorial. Then,
it is natural to assume that the things become easier with the
part of the diffeomorphisms group which does not change $\theta$.
For a non-degenerate $\theta^{\mu\nu}$ these diffeomorphisms
(symplectomorphisms) are generated by vector fields of the form
\begin{equation}
\xi^\mu (x)=\theta^{\mu\nu}\partial_\nu f(x)\,. \label{sym}
\end{equation}
Such diffeomorphisms preserve also the volume element, and thus we
are dealing with unimodular gravity theories. NC theories based on
gauging symplectic diffeomorphisms were indeed constricted
\cite{sdif} and gave rise to many interesting results. Though in
our rather strict approach to NC gravities this group looks too
small, we again receive an important message that a consistent NC
theory may be constructed at least with this small part.

\subsection{Gravity through Yang-Mills type symmetries}
The action of a Yang-Mills gauge transformation can easily be extended
to a noncommutative case. Let in a commutative theory 
$\delta_\alpha \phi = \rho(\alpha) \cdot \phi$, where $\phi$ is 
a field transformed according to a finite dimensional representation 
$\rho$ of the symmetry algebra. Then in an NC case one can define
$\delta_\alpha^\star \phi = \rho(\alpha)\star \phi$. A problem appears 
with commutators. Let $T^A$ be a basis in
the Lie algebra taken in the representation $\rho$. Then
\begin{eqnarray}
&&\delta_\alpha^\star \delta_\beta^\star -
\delta_\beta^\star \delta_\alpha^\star =\delta_{[\alpha,\beta]_\star}^\star
\nonumber\\
&&
[\alpha,\beta]_\star
=\frac 12 [T^A,T^B] (\alpha_A \star \beta_B + \beta_B \star \alpha_A)
\nonumber\\ &&\qquad\qquad +
\frac 12 \{ T^A,T^B\}  (\alpha_A \star \beta_B - \beta_B \star \alpha_A)
\nonumber
\end{eqnarray} 
The expression on the right hand side of the last line is a gauge
generator if both commutator $[T^A,T^B]$ and anticommutator $\{ T^A,T^B\}$
belong to the Lie algebra. This imposes severe restrictions on possible 
gauge groups and their representations\cite{Chaichian:2001mu}.
For example, $su(n)$ cannot be extended to NC spaces, while $u(n)$ can.

One can demonstrate, that with the choice of the potentials
$U(X)=0$, $V(X)\propto X$ corresponding to the Jackiw-Teitelboim
model\cite{JT} is equivalent to an $su(1,1)$ BF theory. Consequently,
extending this symmetry to an NC $u(1,1)$ one can construct an NC
version of the JT gravity\cite{Cacciatori:2002ib}. The model appears
to be both classical\cite{Cacciatori:2002ib} and 
quantum\cite{Vassilevich:2004ym} integrable. Of course, by extending
the gauge symmetry one introduces a new gauge field, which, however, 
decouples in the commutative limit and does not lead to any contradictions.
However, there is a different problem with this approach. One cannot deform
the linear dilaton potential $V(X)$ by adding higher powers of the dilaton
and preserving the number of NC gauge symmetries\cite{Vassilevich:2006uv}.
This means that other interesting dilaton gravity models cannot be
constructed in this approach.
\subsection{Twisted symmetries}
Practically all symmetries of commutative theories can be realized
on a noncommutative space as {\it twisted} symmetries. The twisting
is based on an observation that the Moyal product (\ref{Moyal})
can be represented as a composition of the point-wise product and a
Drinfeld twist. Indeed, the point-wise product $\mu : A\otimes A \to A$,
$\mu (f_1\otimes f_2)=f_1\cdot f_2$ and the Moyal product 
$\mu_\star: A\otimes A \to A$, $\mu_\star (f_1\otimes f_2)=f_1\star_M f_2$
are related through $\mu_\star = \mu \circ \mathcal{F}^{-1}$, where
\begin{equation}
\mathcal{F}=\exp \mathcal{P},\qquad \mathcal{P}=-\frac i2 \theta^{\mu\nu}
\partial_\mu \otimes \partial_\nu \label{twop}
\end{equation}
is a twist.

The way how the symmetry generators act on tensor products is defined
by the coproduct $\Delta$. In commutative field theories one uses a
primitive coproduct 
$\Delta_0(\alpha)=\alpha \otimes 1 + 1 \otimes \alpha$,
so that we have the usual Leibniz rule
\begin{equation}
\alpha (\phi_1 \otimes \phi_2) = \Delta_0(\alpha) (\phi_1 \otimes \phi_2)
=(\alpha \phi_1)\otimes \phi_2 + \phi_1 \otimes (\alpha \phi_2).
\label{prim}
\end{equation}

 We may define another (twisted) coproduct
\begin{equation}
\Delta_{\mathcal{F}}=\mathcal{F}\Delta \mathcal{F}^{-1}\label{twic}
\end{equation}

The action of a generator $\alpha$ on the star-product of fields is defined
as follows
\begin{equation}
\alpha (\phi_1\star_M \phi_2)=\mu_\star (\Delta_{\mathcal{F}}(\alpha) 
\phi_1\otimes\phi_2)=\mu\circ \mathcal{F}^{-1} (\Delta_{\mathcal{F}}(\alpha) 
\phi_1\otimes\phi_2)
\label{alstar}
\end{equation}
Twisting, in a sense, pushes the symmetry generator through the star product.
This makes it possible to define symmetry transformations without
transforming the star product. In algebraic language, we have a
Hopf algebra symmetry instead of a Lie algebra one.

The literature on twisted symmetries is very large. We like to mention
an early paper by Oeckl\cite{Oeckl:2000eg}. The symmetries relevant 
for our discussion are the Poincare symmetry\cite{tPoi} (this was
the first symmetry to be twisted), 
diffeomorphisms\cite{tdif}, and gauge symmetries\cite{tgau}.
Moreover, the twist interpretation may be given to some star products
other than the Moyal one.

Twisting the diffeomorphism transformations allowed to define a model
of NC gravity\cite{tdif} invariant under the full diffeomorphism algebra,
though this invariance is realized in a non-standard way\footnote{
There are also critics of twisting local symmetries, see
\cite{Chaichian:2006wt}.}.

The twisted symmetries are not \emph{bona fide} physical symmetries.
One cannot use them, for example, to gauge away any degrees of freedom.
The problem of proper interpretation of twisted local symmetries
remains. One possible interpretation is as follows \cite{Vassilevich:2007jg}.
Let us replace the partial derivatives $\partial$ in (\ref{Moyal})
and (\ref{twop})
with covariant derivatives $\nabla$ with a trivial connection.
Since $\nabla_\mu$ commute, the new star product will be again
associative. (For non-commuting $\nabla$ the associativity is violated
\cite{Riv}).
If the original theory were twisted gauge invariant,
the theory with this new star product will be both twisted gauge 
invariant and gauge invariant in the ordinary sense. To return back, 
one has to fix the gauge $\nabla = \partial$. Therefore, twisted
gauge invariance is a remnant of ordinary gauge invariance after
fixing the gauge by imposing a condition on gauge-trivial covariant
derivatives appearing inside the star product.

\subsection{NC geometry and spectral action}
A unifying approach to describe \emph{any} NC geometry was introduced
by Connes\cite{Connes} (see also Ref.\cite{MullerHoissen:2007xy} for 
a recent overview). It is based on the notion of a spectral triple
$(A,H,D)$ consisting of an associative algebra $A$ represented by
bounded operators on a Hilbert space $H$ and a Dirac operator $D$
acting on $H$. These three object satisfy certain relations and
restrictions. As soon as a spectral triple is defined, the corresponding
classical action follows from the so-called spectral 
action principle\cite{sap} 
\begin{equation}
S={\mathrm{Tr}}\, \Phi (D/\Lambda),\label{spec}
\end{equation}
where $\Phi$ is a positive even function, and $\Lambda$ is a scale parameter.
All unitary symmetries of the operator $D$ are inherited by the spectral
action. As an expansion in $\Lambda$ the action (\ref{spec}) may be
calculated by the heat kernel methods. On Moyal spaces such methods are
rather well developed \cite{Vassilevich:2007fq}. The problem is ``only''
to find a corresponding spectral triple.  

A similar idea, that the NC gravity may be induced is explored in the
emergent gravity approach, see Ref.\cite{Steinacker:2007dq} and references
therein.
\section{The star products}
As we have seen above, rigidity of $\theta^{\mu\nu}$ under the
diffeomorphism transformations creates a lot of problems. It may 
be a good idea to transform both $\theta^{\mu\nu}$ and the
star product under the diffeomorphisms. To this end, we need
general star products.

The modern history of deformation quantization started with the 
papers \cite{BFFLS}. The main part of the deformation-quantization 
program is a construction of a star product for a given Poisson
structure $\theta^{\mu\nu}(x)$. For symplectic manifolds (non-degenerate
$\theta^{\mu\nu}$) the existence of a start product was demonstrated
by De Wilde and Lecomte\cite{DWL}, and a very elegant construction
was given by Fedosov\cite{Fedosov}. For generic Poisson structure the
existence of a star product was demonstrated by Kontsevich 
\cite{Kontsevich:1997vb} who also gave an explicit formula (which is,
however, too complicated to be used for actual calculations of 
higher orders in the
star product). Such orders of the star product were computed by using
the Weyl map and a representation of noncommutative coordinates in the
form of differential operators\cite{Kupriyanov:2008dn}.

A very promising non-perturbative formula for the star product was
suggested by Cattaneo and Felder \cite{Cattaneo:1999fm}.
They took a Poisson sigma model with the action
\begin{equation}
S_{\rm PSM}=\int \left[ A_\mu dX^\mu + \frac 12
\theta^{\mu\nu}(X)A_\mu \wedge A_\nu \right] \label{PSM}
\end{equation}
defined on a two-dimensional manifold. $X$ and $A$ are the fields
on this manifold, which are a zero-form taking values in a Poisson
manifold  and a one-form with values in the cotangent space to
this manifold, respectively. The two-dimensional world-sheet is
supposed to be a disc (with suitable boundary conditions imposed
on $A$). Three distinct points on the boundary of the disc are
selected, denoted $0$, $1$, and $\infty$. The star product is then
given by a correlation function
\begin{equation}
f\star g(x)=\int dA\, dX\, f(X(0))g(X(1))\, e^{iS_{\rm PSM}}\,,
\label{CFstar}
\end{equation}
where the integration is restricted by the condition
$X^\mu(\infty)=x^\mu$. The main advantage of this formula is that
it does not imply any expansion in $\theta$.

What is then the relation to two-dimensional dilaton gravities?
The point is that the Poisson sigma models were originally
introduced\cite{Schaller:1994es,Ikeda:1993fh}
as generalizations of the dilaton gravity action
(\ref{1st}). Indeed, by identifying $X, X^a$ with $X^\mu$, and $\omega,
e^a$ with $A_\mu$ and making a suitable choice of
$\theta^{\mu\nu}(X)$ one can reduce (\ref{PSM}) to (\ref{1st}).
In the context of two-dimensional gravities rather powerful
methods of calculation of the path integral were developed\cite{KLV}.
At least some of these methods work also for generic Poisson sigma
models\cite{Hirshfeld:1999xm}. The approach\cite{KLV} was specially
tailored to study quantum gravity phenomena, like virtual black holes,
and not the correlation functions of the type (\ref{CFstar}). However,
some steps to adjust that methods to the new tasks have already been
done. For example, inclusion of boundaries was 
considered in a paper\cite{Bergamin:2005pg}, which was the last 
publication of Wolfgang Kummer.

\section{Conclusions}
As we have seen, there are many rather successful approaches to
NC gravity. One can be optimistic, that soon an NC gravity satisfying
our (perhaps, too strict) criteria will be formulated. It is likely,
that 2D dilaton gravities will play a prominent role in this process.
\section*{Acknowledgements}
This work was supported in part by CNPq (Brazil).

\end{document}